# Antiferromagnetic to Ferrimagnetic Phase Transition and Possible Phase Coexistence in Polar Magnets $(Fe_{1-x}Mn_x)_2Mo_3O_8$ ($0 \leq x \leq 1$)

*Yuting Chang[1#], Lei Gao[1#], Yunlong Xie[2], Bin You[1], Yong Liu[3], Rui Xiong[3], Junfeng Wang[1], Chengliang Lu[1\*], and Jun-Ming Liu[2,4]*

[1] School of Physics & Wuhan National High Magnetic Field Center, Huazhong University of Science and Technology, Wuhan 430074, China

[2] Institute for Advanced Materials, Hubei Normal University, Huangshi 435001, China

[3] School of Physics and Technology, and the Key Laboratory of Artificial Micro/Nano structures of Ministry of Education, Wuhan University, Wuhan 430072, China

[4] Laboratory of Solid State Microstructures and Innovation Center of Advanced Microstructures, Nanjing University, Nanjing 210093, China

**Abstract:**

In the present work, magnetic properties of single crystal $(Fe_{1-x}Mn_x)_2Mo_3O_8$ ($0 \leq x \leq 1$) have been studied by performing extensive measurements. A detailed magnetic phase diagram is built up, in which antiferromagnetic state dominates for $x \leq 0.25$ and ferrimagnetic phase arises for $x \geq 0.3$. Meanwhile, sizeable electric polarization of spin origin is commonly observed in all samples, no matter what the magnetic state is. For the samples hosting a ferrimagnetic state, square-like magnetic hysteresis loops are revealed, while the remnant magnetization and coercive field can be tuned drastically by simply varying the Mn-content or temperature. Possible coexistence of the antiferromagnetic and ferrimagnetic phases is proposed to be responsible for the remarkable modulation of magnetic properties in the samples.

# I. Introduction

Magnetoelectrics has been providing a fertile ground for studying cross-coupling among various ferroic-orders, which is fundamentally interesting and practically important for developing conceptually new devices [1-4]. Recently, a group of newly discovered magnetoelectric (ME) materials, i.e. the polar magnets $A_2Mo_3O_8$ (*A*: transition metal), have been drawing considerable attention, because of their concurrent spontaneous electric polarization (*P*) due to the crystallographic polarity, spin-driven electric *P*, and linear ME effect [5-11]. This is rarely seen in the existing materials such as other ferrites [12-13]. Note that the linear magnetoelectrics usually have zero spontaneous *P* and thus are regarded as non-ferroelectric (FE) [14].

The family of $A_2Mo_3O_8$ belongs to a pyroelectric space group $P6_3mc$ [15-17]. As illustrated in Fig. 1(a), there are two inequivalent sites of the *A* ions, i.e. the octahedra and the tetrahedra, and the polar arrangement of the $AO_4$ tetrahedra induces crystallographic polarity along the *c*-axis. Magnetism of $A_2Mo_3O_8$ is dominated by the *A*-ions, while the Mo ions are trimerized. The two magnetic sublattices composed of tetragonal-*A* and octahedral-*A* are coupled in an antiferromagnetic (AFM) manner. In each of the sublattices, magnetic moments can be either parallel or antiparallel, depending on the *A* ions. For instance, $Fe_2Mo_3O_8$ and $Co_2Mo_3O_8$ have two interacted AFM sublattices, and $Mn_2Mo_3O_8$ hosts a ferrimagnetic (FRM) ground state where two types of ferromagnetic (FM) orders are coupled antiferromagnetically [9, 18]. In $Ni_2Mo_3O_8$, a specific noncollinear magnetic configuration is stabilized due to the coupling of magnetic sublattices of $Ni^{2+}$ [11]. The different magnetic configurations give rise to strikingly distinct ME effect and thereby underlying physics in these materials.

Beyond the conventional cross-control between magnetic and electric variables, some more fascinating properties such as the optical ME effect which is an extension of the classic ME to dynamic region, orbitally selective Mott feature, and giant thermal Hall effect have been revealed typically in $Fe_2Mo_3O_8$. Importantly, all these properties are found to closely relate to the magnetic structure. The band gap of $Fe_2Mo_3O_8$ exhibits notable narrow down at the AFM transition $T_N \sim 60$ K, implying intimate coupling between the charge and spin degrees of freedom [19]. The optical ME depends on the relative arrangement between magnetization ($M$) and electric polarization $P$, e.g. the diagonal term $M \cdot P$ leads to the gyrotropic birefringence and the off-diagonal term $M \times P$ determines the nonreciprocal directional dichroism [20-21]. For the thermal Hall effect, the transverse thermal conductivity jumps to a unprecedentedly large value driven by the AFM to FRM transition [7]. Motivated by these observations, it is of highly interesting and practical to study the modulation of magnetic order of $Fe_2Mo_3O_8$, which may extend the ME scenario to a much broader scope.

$Fe_2Mo_3O_8$ hosts a rather robust AFM ground state, and the critical magnetic field ($H_{cri}$) of the AFM-FRM transition is quite high indeed [5, 8]. In this sense, chemical substitution may be a more favorable route to tailor the magnetism of $Fe_2Mo_3O_8$, considering the very different magnetic states among the members of $A_2Mo_3O_8$ as mentioned above. In particular, it was found that the Fe ions at the octahedral ($Fe_O$) and the tetrahedral ($Fe_T$) sites could be selectively substituted by properly choosing the doping species [22]. This is fundamentally important, because of the distinct electronic and magnetic properties between $Fe_O$ and $Fe_T$. In comparison with non-magnetic ions, doping with magnetic ions could better preserve the fascinating properties that essentially associate with magnetic order. Especially, the effect of Mn-substitution at Fe-site on the magnetic properties deserves special attention, based on three motivations. First, $Fe_2Mo_3O_8$ hosts an AFM ground state, while $Mn_2Mo_3O_8$ has a FRM ground state, which facilitates the modulation of magnetic structure efficiently. Second, although both materials have the same *c*-axis magnetic anisotropy, $Fe^{2+}$ has much stronger spin anisotropy than $Mn^{2+}$. For instance, the *c*-axis magnetic anisotropy remains notable up to room temperature (*T*) in $Fe_2Mo_3O_8$, but disappears right above the FRM temperature $T_C$~41 K in $Mn_2Mo_3O_8$ [6, 8]. Third, in $Fe_2Mo_3O_8$, $Fe_T$ (~4.6 $\mu_B$/Fe) has much larger magnetic moment than $Fe_O$ (~4.2 $\mu_B$/Fe) [22]. Differently, $Mn_O$ and $Mn_T$ have the same magnetic moments of ~5 $\mu_B$/Mn in $Mn_2Mo_3O_8$, which are expected to fully compensate at *T*=0 K. Moreover, it was found that Mn-doping in $Fe_2Mo_3O_8$ indeed causes interesting sign-reversal of the linear ME coefficient from negative to positive, while the detailed magnetic properties of $(Fe_{1-x}Mn_x)_2Mo_3O_8$ remains largely unexplored [6].

In the present work, a series of single crystal $(Fe_{1-x}Mn_x)_2Mo_3O_8$ (0≤*x*≤1) were synthesized, and extensive characterizations including structure, magnetization, and pyroelectric current measurements were carried out. Doping induced AFM to FRM transition is identified at *x*=0.25, i.e. the bicritical point. It is found that the FRM phase can be modulated remarkably, and the remnant magnetization ($M_r$) and coercive field ($H_c$) exhibit strong dependence on the Mn-content *x* and *T*. Interestingly, possible coexistence of AFM and FRM phases may widely exist in the materials with different *x*.

## II. Experimental details

The $(Fe_{1-x}Mn_x)_2Mo_3O_8$ single crystals were synthesized using chemical vapor transport method [23-24]. High purity $MnO_2$, Mn, $Fe_2O_3$, CoO, and $MoO_2$ powders were mixed and sealed in evacuated silica tubes, which then were placed in a two-zone furnace for the crystal growth. Chemical vapor transport method is more favorable for the synthesis of $(Fe_{1-x}Mn_x)_2Mo_3O_8$ polar magnets than other routes [25-26]. The obtained crystals are hexagonal plate-like with typical size of $2\times2\times0.5$ mm$^3$. In order to determine the crystalline structure, the crystals were crashed thoroughly, and then powder X-ray diffraction (XRD) measurements were performed at room temperature ($T$). Energy Dispersive X-ray Spectroscopy (EDS) and X-ray Photoelectron Spectroscopy (XPS) measurements were performed to check the stoichiometry and valance state of the single crystals. The magnetization measurements as a function of $T$ and magnetic field $H$ were carried out using a Quantum Design physical property measurement system (PPMS). For both zero field cooling (ZFC) and field cooling (FC) measurements, the measuring field was fixed at $H=0.1$ T. The $M(H)$ measurements were carried out at various temperatures for the samples. All the magnetization measurements were performed with $H//c$-axis. Gold electrodes were prepared for pyroelectric current measurements. Electric polarization along the $c$-axis as a function of $T$ were obtained by integrating the pyroelectric current which were measured using Keithely 6517 electrometer during the warming process. Pulsed high magnetic field measurements for magnetization were performed by means of a standard inductive method employing a couple of coaxial pickup coils. The pulsed magnetic field up to ~58 T with a duration time of ~10 ms was generated by using a nondestructive long-pulse magnet energized by a 1 MJ capacitor bank.

## III. Results and Discussions

The collected powder XRD spectra confirm that all samples with different Mn-content $x$ are pure phase. To acquire more details of the crystalline structure, Rietveld-profile refinements of the XRD data were performed for all samples. The refined result for the sample with $x=0.9$ is shown as an example in Fig. 1(b). The difference between the measured and refined spectra is small with reliability parameter $R_{wp}$ = 3.86%. For all other samples, the refinements are all high quality, and the $R_{wp}$ values are at similar levels. According to the refinements, derived lattice constants are $a$=5.773 Å and $c$=10.054 Å for $Fe_2Mo_3O_8$ (i.e. $x$=0), and $a$=5.799 Å and $c$=10.267 Å for $Mn_2Mo_3O_8$ (i.e. $x$=1), which are in good agreement with previous studies [18]. The obtained $a$, $c$, and cell volume $V$ are plotted as a function of $x$ in Fig. 1(c)-(e), respectively. It is seen that all lattice parameters vary with $x$ continuously, without showing any apparent anomaly. Since stoichiometry and valance state are crucial ingredients in determining the properties of oxides [27-30], further EDS and XPS measurements have been performed in the crystals (shown in the Supporting Information). It is confirmed that no apparent deviation of the concentration of the original cations from a given value, and $Fe^{2+}$ and $Mn^{2+}$ are quite stable in the crystals.

Because of the inequivalent sites, $Fe_2Mo_3O_8$ has two AFM sublattices consisting of $Fe_O$ and $Fe_T$, respectively, which are coupled antiferromagnetically. $Fe_2Mo_3O_8$ undergoes a paramagnetic (PM) to AFM transition at $T_N$~60 K, evidenced by sharp peaks in $M_c(T)$ curves measured under both ZFC and FC sequences, as shown in Fig. 2(a). By introducing $Mn^{2+}$ into the $Fe^{2+}$ sites, $T_N$ is gradually shifted toward low-$T$ region. For the sample with $x$=0.2 shown in Fig. 2(c), the peak of $M_c(T)$ gets evidently higher than the samples with lower doping content, indicating the occurrence of ferrimagnetism in the sample. At $x$=0.25, the $M_c(T)$ curves exhibit drastic enhancement ensuing at $T_C$~53 K, and the subsequent peak is about ten times higher than that of the samples with $x$<0.2, evidencing development of the FRM phase. The FRM transition temperature $T_C$ is determined by the minima of $T$ derivative of $M_c$. Further cooling the sample causes steep decrease in $M_c$ at $T_N$~39

K, owing to the FRM to AFM transition.

At $x=0.3$, the FRM phase is well stabilized below $T_C\sim53$ K, and no further AFM transition can be seen in the low-$T$ region according to the FC curve. In ferrimagnets, there are generally two ways to induce net $M$ [31]. One is due to the different magnetic moments of the interacted sublattices, and the uncompensated $M$ emerges below $T_C$ and then persists down to low $T$. The other way is related to different $T$-dependence of the sublattices hosting the same magnetic moment. In this sense, the $M(T)$ curve usually first increases fast at $T_C$, and then gradually decreases with $T$ upon cooling the system. Eventually, $M$ goes to zero as $T\rightarrow0$, because of cancellation of the two sublattices. For a mixed case where the sublattices have different $T$-dependence and magnetic moments, $M$ is expected to be reduced to a finite value at low-$T$, such as the FC curve at $x=0.3$ shown in Fig. 2(e). Another phenomenon is that the ZFC curve shows tremendous difference from the FC curve and becomes negative below $T_B\sim20$ K. This has been widely observed in other ferrimagnets such as spinels, and can be understood based on the classic molecular field theory [32]. The two FM sublattices are aligned in opposite directions, and the sign of net $M$ depends on the competition between Zeeman energy and magnetic anisotropy. As a consequence, negative $M$ can appear. Once the Zeeman energy is beyond the anisotropy energy, the net $M$ is thus reversed, leading to the sharp variation of $M_c$ from negative to positive around $T_B\sim20$ K in the ZFC curve.

Further increasing $x$ causes continuous decrease in $T_C$. Meanwhile, the large bifurcation between ZFC and FC curves can be seen in all samples with $x>0.3$, and $T_B$ is changed modestly with $x$. Obviously, the FC curves of the samples commonly exhibit non-zero $M_c$ as $T\rightarrow0$ K, resulting from the different magnetic moments of the sublattices. In $Mn_2Mo_3O_8$ ($x=1$), the PM-FRM transition is identified at $T_C\sim41$ K, consistent with previous study. Although $Mn_O$ and $Mn_T$ have the same magnetic moment of 5 $\mu_B$/Mn, the different $T$-dependence of two FM sublattices leads to the striking $M_c$-enhancement at $T_C$, a typical example of ferrimagnetism as mentioned above. Nevertheless, a fully compensated state would be eventually achieved at $T$=0 K, revealed by the fast approaching between the FC and ZFC curves below $T_B$~12 K shown in Fig. 2(m).

To better understand the magnetic properties of $(Fe_{1-x}Mn_x)_2Mo_3O_8$, extensive $M(H)$ measurements were carried out, and typical results are shown in Fig. 3. For the non-doped sample $Fe_2Mo_3O_8$, sharp AFM to FRM transition is triggered by relatively large critical field $H_{cri}$, e.g. $H_{cri}$~12 T at $T$=35 K. The jump at $H_{cri}$ in $M(H)$ is caused by the different magnetic moments between $Fe_T$ and $Fe_O$, which has similar value of about ~0.5 $\mu_B$/f.u. at various temperatures. Increasing $x$ causes significant reduction in $H_{cri}$. For instance, $H_{cri}$ is decreased from ~12 T to ~3 T at $T$= 35 K as $x$ is increased from 0 to 0.2, indicating strong suppression of the AFM phase of the system. At $x$=0.2, additional kink anomaly arises in $M(H)$ above $H_{cri}$. This is getting more clear at $x$=0.25 where multiple step-like transitions accompanied by evident magnetic hysteresis can be identified, shown in Fig. 3(d). In fact, weak but unambiguous magnetic hysteresis also exists at the very low-$H$ region. These phenomena imply coexistence of the AFM and FRM phases in the material, and the steps are due to successive transitions from AFM to FRM domains, consistent with the observation of both $T_C$ and $T_N$ from the $M(T)$ shown in Fig. 2(d).

For the sample with $x$=0.3 shown in Fig. 3(e), square-like $M(H)$ hysteresis loops are obtained, confirming the emergence of the FRM phase. Interestingly, the initial branch of $M(H)$ at $T$=15 K shows step-like transition at a reduced critical field $H$~0.5 T, possibly due to the $H$-driven AFM-FRM transition akin to the cases at $x$≤0.25. Similar phenomenon is generally seen in the samples with higher doping content. Meanwhile, the doped samples with 0.3≤$x$≤0.9 exhibit very large coercive field ($H_c$) shown in Fig. 3(e)-(k), which is about one order of magnitude larger than that of $Mn_2Mo_3O_8$ hosting a pure FRM state (Fig. 3(l)). Therefore, strong competition between AFM and FRM phases may be a common feature in the $(Fe_{1-x}Mn_x)_2Mo_3O_8$ samples with intermediate doping levels.

In addition, the $M(H)$ curve at $T$=15 K of $Mn_2Mo_3O_8$ displays another clear upturn above $H$~5 T, shown in Fig. 3(l). To understand this, high field magnetization measurements were carried out, which are able to capture much more details of the sudden upturn in a broad $H$-range. As shown in Fig. 4(a), the critical field of the upturn is moved to higher $H$ side with increasing $T$. This is basically different from conventional FRM to FM transition which is usually promoted by thermal fluctuation, e.g. the critical field would be lowered by increasing $T$ of the sample. Moreover, $M(H)$ shows continuous linear increase after the upturn, without any trace of saturation. This resembles the typical antiferromagnetic behavior. Therefore, the upturn-anomaly should be assigned to a spin-flop transition from the $c$-axis FRM to the $ab$-plane FRM. The critical field of the upturn at various temperatures is summarized in Fig. 4(b).

Remnant magnetization $M_r$ and coercive field $H_c$ are two key parameters of magnetic materials. In fact, enhancing the macroscopic magnetization has been one of the primary goals of the ME research, which is crucial to combine the advantages of both magnetic and ferroelectric memories. Typical data of $M_r$ and $H_c$ derived from $M(H)$ at various temperatures are plotted as a function of Mn-content $x$ in Fig. 5(a) and (b), respectively. Similar behavior is observed for $M_r(x)$ and also $H_c(x)$ at different $T$. As an example, at $T$=15 K, $M_r$ jumps from zero to ~0.2 $\mu_B$/f.u. at $x$=0.3 where the doping-induced FRM phase is stabilized, and then decreases gradually for 0.4≤$x$≤1. $H_c(x)$ displays different evolving way that it increases quickly as the FRM phase is induced, and then reaches a plateau for 0.5≤$x$≤0.9. After that, $H_c$ drops to a small value of ~0.2 T in $Mn_2Mo_3O_8$, i.e. $x$=1. The full data of both $M_r$ and $H_c$ at various $T$ and $x$ are summarized in Fig. 5(c) and (d), respectively. $M_r$ increases with $T$ monotonously for all samples, and it can be as large as ~0.5 $\mu_B$/f.u. for the samples with 0.3≤$x$≤0.5. Regarding $H_c(T)$, successive decrease can be seen for all samples hosting the FRM phase.

The above magnetization characterizations have demonstrated highly tunable magnetic properties in $(Fe_{1-x}Mn_x)_2Mo_3O_8$. Further, it is confirmed that the magnetically driven electric polarization is maintained in all samples, no matter what kind of magnetic state is hosted. Each sample was first cooled down far above $T_N$ or $T_C$, and then the pyroelectric current was collected during the warming process. For both the cooling and warming processes, no magnetic and electric fields were applied. Because of the polar crystalline structure, all samples display notable pyroelectric current even far above $T_N$. Importantly, additional sharp valley emerges in each pyroelectric current curve, which matches well with the AFM or FRM transition of the sample, shown in Fig. 6(a). By integrating the pyroelectric current versus time, $T$-dependence of electric polarization is obtained, shown in Fig. 6(b). The marked drop of Δ$P(T)$ at $T_C$ or $T_N$ suggests a spin origin of the induced $P$, probably due to an exchange striction mechanism proposed by previous first principles calculations [8]. Moreover, magnitude of the $P$-drop is about Δ$P$ ~1000 μC/m$^2$ for the samples, which is larger than most of the multiferroics hosting a spin-origin [2, 33-37].

Based on the comprehensive characterizations of magnetization and pyroelectric current, a $T$-$x$ phase diagram is built up for $(Fe_{1-x}Mn_x)_2Mo_3O_8$ shown in Fig. 7(a). Three regions can be seen in the diagram, which are dominated by PM, FRM, and AFM phases, respectively. At the phase boundary $x\sim0.25$, coexistence of AFM and FRM phases is identified. Interestingly, such phase coexisting feature may widely happen in the $(Fe_{1-x}Mn_x)_2Mo_3O_8$ samples, especially for the ones with middle doping levels. Moreover, for the FRM phase, very large remnant magnetization $M_r$ and coercive field $H_c$ are identified, and both parameters can be tuned remarkably by varying the Mn-content $x$ and $T$. To understand these phenomena, the occupation ratios of $Mn^{2+}$ at the two inequivalent sites could be a pivotal factor, because of the very different properties of ions at the two sites in the polar magnets.

In the FRM phase, $M_r$ is in principle determined by the difference of magnetic moments between $Fe_T$ ($Mn_T$) and $Fe_O$ ($Mn_O$), and thus can be given as:

$$M_r = 2x \cdot P_O \cdot M_{Mn-O} + (1 - 2x \cdot P_O) \cdot M_{Fe-O} - [2x \cdot P_T \cdot M_{Mn-T} + (1 - 2x \cdot P_T) \cdot M_{Fe-T}] \quad (1)$$

where $P_O$ and $P_T$ (=1- $P_O$) are the occupation ratios of $Mn^{2+}$ locating at the octahedral site and tetragonal site, respectively. $M_{Fe-O}$ and $M_{Mn-O}$ ($M_{Fe-T}$ and $M_{Mn-T}$) are the magnetic moment of $Fe^{2+}$ and $Mn^{2+}$ on the octahedral (tetragonal) site, respectively. Equation (1) can also be written as:

$$M_r = 2x \cdot P_O \cdot (M_{Mn-O} - M_{Fe-O}) - [2x \cdot P_T \cdot (M_{Mn-T} - M_{Fe-T})] + (M_{Fe-O} - M_{Fe-T}) \quad (2)$$

As revealed by previous Möessbauer spectroscopy on $Fe_2Mo_3O_8$, $M_{Fe-O}$ and $M_{Fe-T}$ are ~4.8 $\mu_B$ and ~4.2 $\mu_B$, respectively [22], resulting in ($M_{Fe-O} - M_{Fe-T}$)~0.6 $\mu_B$. This is close to our experimental result of ~0.5 $\mu_B$, determined by the jump in $M(H)$ due to the AFM-FRM transition. For $Mn_2Mo_3O_8$, $M_{Mn-O}$ and $M_{Mn-T}$ have the same magnitude of $M\sim5$ $\mu_B$. With these parameters, it is now able to simulate the $M_r(x)$. The $M_r(x)$ data typically measured at $T$=40 K close to $T_C$ (Fig. 5(a)) are used for the simulation, in order to make sure that the majority of the domains have been converted to be FRM and uniformly aligned. The obtained occupation ratios of $Mn^{2+}$ are roughly distributed around $P_T \sim 0.6$ and $P_O \sim 0.4$ in all samples, as shown in Fig. 7(b). This is consistent with previous Möessbauer spectroscopy characterizations giving that $Mn^{2+}$ has slight preference to occupy the tetragonal sites in $(FeMn)_2Mo_3O_8$ [6, 18]. For $Mn_2Mo_3O_8$ ($x=1$), the occupation ratios slightly deviate from 0.5, which is caused by the different $T$-dependence of $M_{Mn}$ at different sites.

At low-$T$, the difference between $M_{Mn-O}$ and $M_{Mn-T}$ becomes negligible, and thus $M_r$ is dominated by the term ($M_{Fe-O} - M_{Fe-T}$). For instance, if all domains are FRM, the ideal low-$T$ $M_r$ is ~0.35 $\mu_B$/f.u in the sample with $x=0.3$, by taking ($M_{Fe-O} - M_{Fe-T}$)~0.5 $\mu_B$ and the derived occupation ratios of $Mn^{2+}$. This is about two times larger than the measured value $M_r$~0.18 $\mu_B$/f.u at $T=10$ K. The discrepancy can be even remarkable at lower $T$, because the measured $M_r$ decreases with $T$ linearly, shown in Fig. 5(c). Therefore, it is naturally expected that AFM orders coexist with the FRM phase in the sample with $x=0.3$, responsible for the significantly reduced $M_r$. Difference between the ideal and measured $M_r$ can also be found in the samples with higher doping contents, while the discrepancy is gradually descended. Taking this scenario that AFM and FRM phases coexist, the giant coercive field $H_c$ can be well understood by the strong phase competition in the system. In addition, the polar magnets $(Fe_{1-x}Mn_x)_2Mo_3O_8$ exhibit large remnant magnetization and electric polarization simultaneously, which is usually accessible in composite ME systems rather than the single-phase materials [38-39], and therefore may have interesting potentials in developing microwave materials [40-41].

## IV. Conclusion

A series of single crystal $(Fe_{1-x}Mn_x)_2Mo_3O_8$ ($0 \leq x \leq 1$) have been synthesized, and characterized by performing structure, magnetization, and electric polarization measurements. It is found that Mn-substitution of Fe can induce antiferromagnetic to ferrimagnetic phase transition at $x=0.25$, and coexistence of AFM and FRM phases may be a common feature in the materials. For the samples with $x \geq 0.3$ where ferrimagnetic state is stabilized, square-like magnetic hysteresis loops with highly tunable remnant magnetization $M_r$ and coercive field $H_c$ are identified. In particular, remarkable electric polarization driven by magnetic ordering is found in all samples hosting various magnetic states. These findings indicate that $(Fe_{1-x}Mn_x)_2Mo_3O_8$ possessing highly tunable properties could be unique candidate for exploring emergent physics and functions.

## ■ ASSOCIATED CONTENT

**Supporting Information**

The Supporting Information is available free of charge at ***

Additional characterizations such as energy dispersive X-ray spectroscopy and X-ray photoelectron spectroscopy of the materials supplied as Supporting Information.

## ■ AUTHOR INFORMATION

**Corresponding Author**

* Authors to whom correspondence should be addressed, email: <cllu@hust.edu.cn>;

**Author Contributions**

# Yuting Chang and Lei Gao contributed equally to this work.

C.L.L. designed and supervised the experiments. Y.T.C. performed the measurements. L.G., B.Y., Y.L., R.X., and J.F.W. contributed to the measurements and data analysis. Y.L.X. contributed to data analysis. C.L.L. and J.M.L. wrote the manuscript.

**Notes**

The authors declare no competing financial interest.

## ■ ACKNOWLEDGMENT

This work is supported by the National Nature Science Foundation of China (Grant Nos. 12174128, 12074291, 92163210, and 11834002), Hubei Province Natural Science Foundation of China (Grant No. 2020CFA083), and the Fundamental Research Funds for the Central Universities (Grant No. 2019kfyRCPY081, 2019kfyXKJC010).

**Figures:**

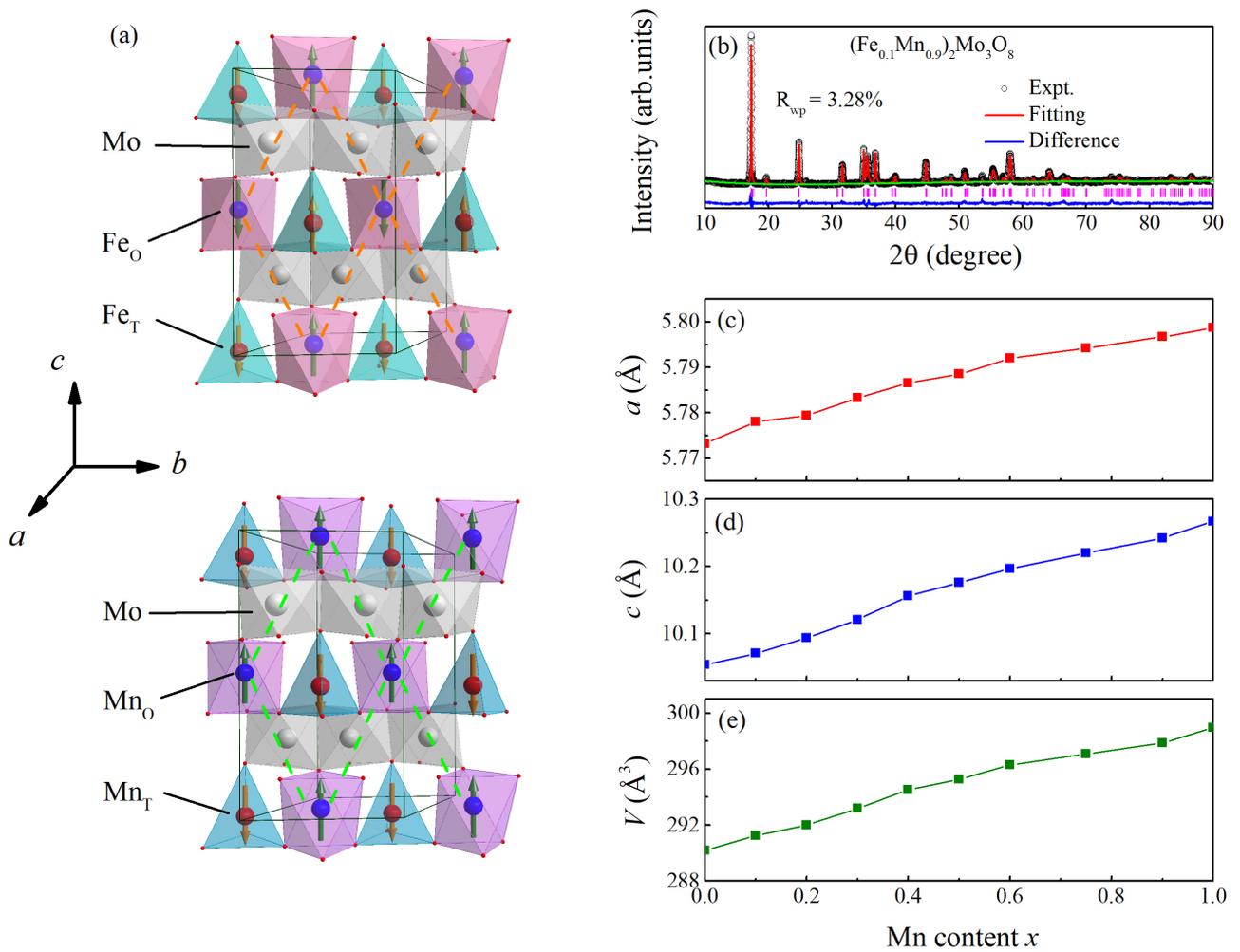

Figure 1. (a) Sketch of crystalline structure of $Fe_2Mo_3O_8$ (top) or $Mn_2Mo_3O_8$ (bottom). Magnetic moments of $Fe^{2+}$ and $Mn^{2+}$ have also been labeled correspondingly. One of the magnetic sublattices is indicated by dashed lines. (b) Rietveld refinement of the XRD spectra of sample $(Fe_{0.1}Mn_{0.9})_2Mo_3O_8$. (c)-(e) present evaluated lattice parameters $a$, $c$, and cell volume $V$ as a function of Mn-content $x$, respectively.

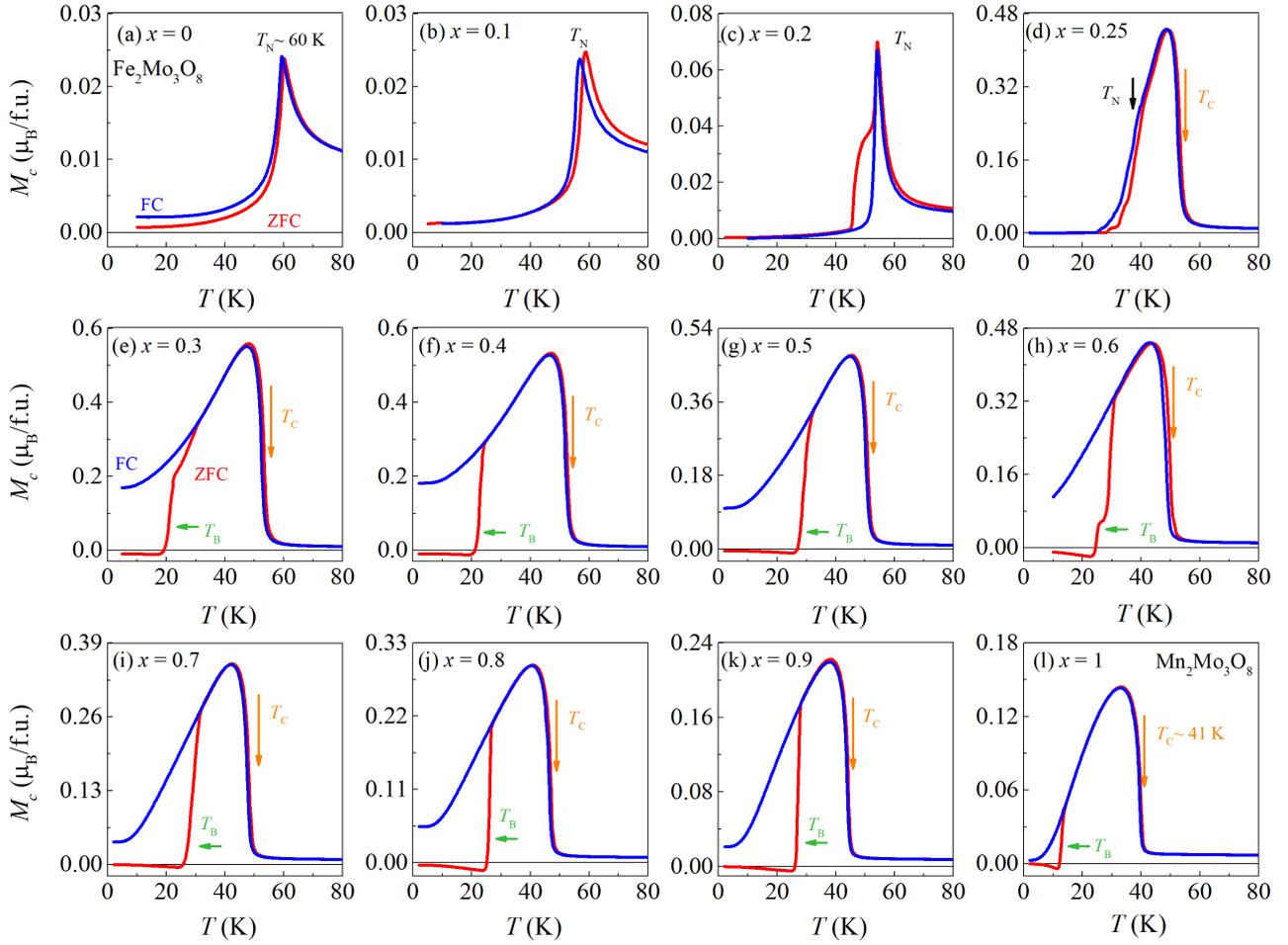

Figure 2. Temperature dependence of magnetization with $H$ along the $c$-axis for $(Fe_{1-x}Mn_x)_2Mo_3O_8$ ($0 \leq x \leq 1$) single crystals: (a) $x=0$, (b) $x=0.1$, (c) $x=0.2$, (d) $x=0.25$, (e) $x=0.3$, (f) $x=0.4$, (g) $x=0.5$, (h) $x=0.6$, (i) $x=0.7$, (j) $x=0.8$, (k) $x=0.9$, and (l) $x=1$. $T_C$ and $T_N$ represent the ferrimagnetic and antiferromagnetic transitions, respectively. $T_B$ indicates the reversal of net magnetization.

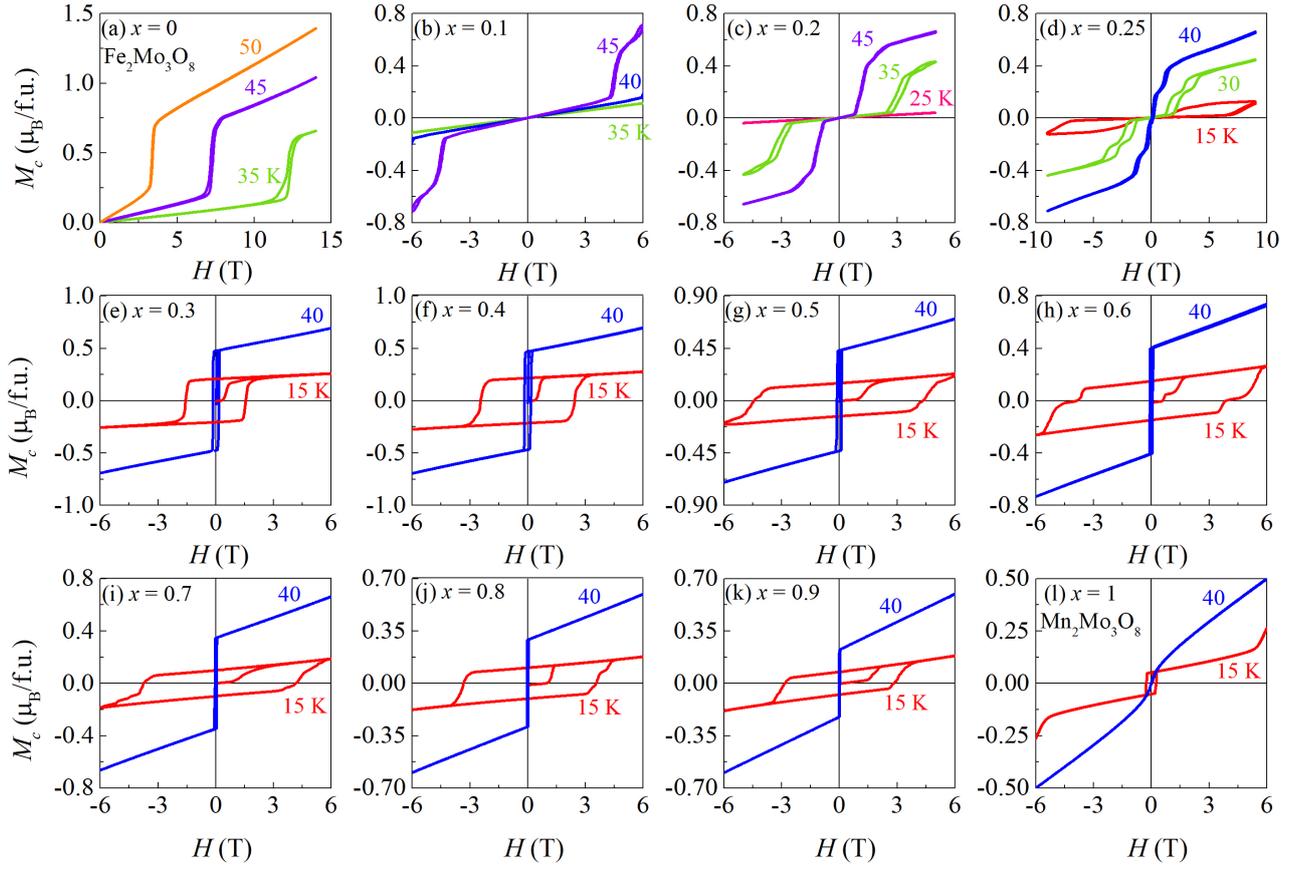

Figure 3. Magnetization as a function of $H \parallel c$-axis measured at various temperatures for $(Fe_{1-x}Mn_x)_2Mo_3O_8$ ($0 \leq x \leq 1$) single crystals: (a) $x=0$, (b) $x=0.1$, (c) $x=0.2$, (d) $x=0.25$, (e) $x=0.3$, (f) $x=0.4$, (g) $x=0.5$, (h) $x=0.6$, (i) $x=0.7$, (j) $x=0.8$, (k) $x=0.9$, and (l) $x=1$.

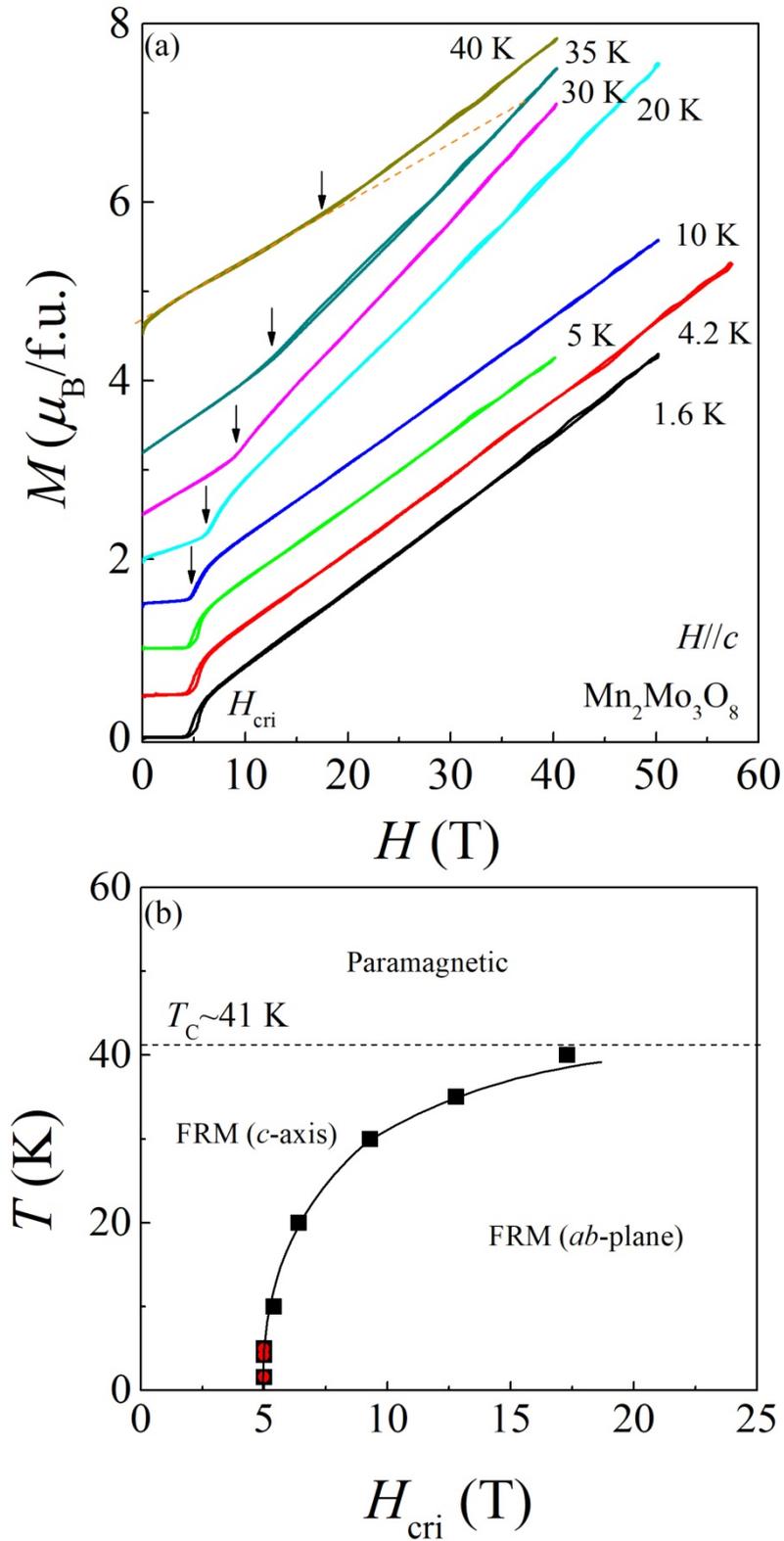

Figure 4. (a) High field magnetization measurements of $Mn_2Mo_3O_8$, in which the critical field $H_{cri}$ of spin flop transition is indicated. (b) Summarized magnetic phase diagram, based on the pulsed (black squares) and static (red dots) magnetic fields measurements.

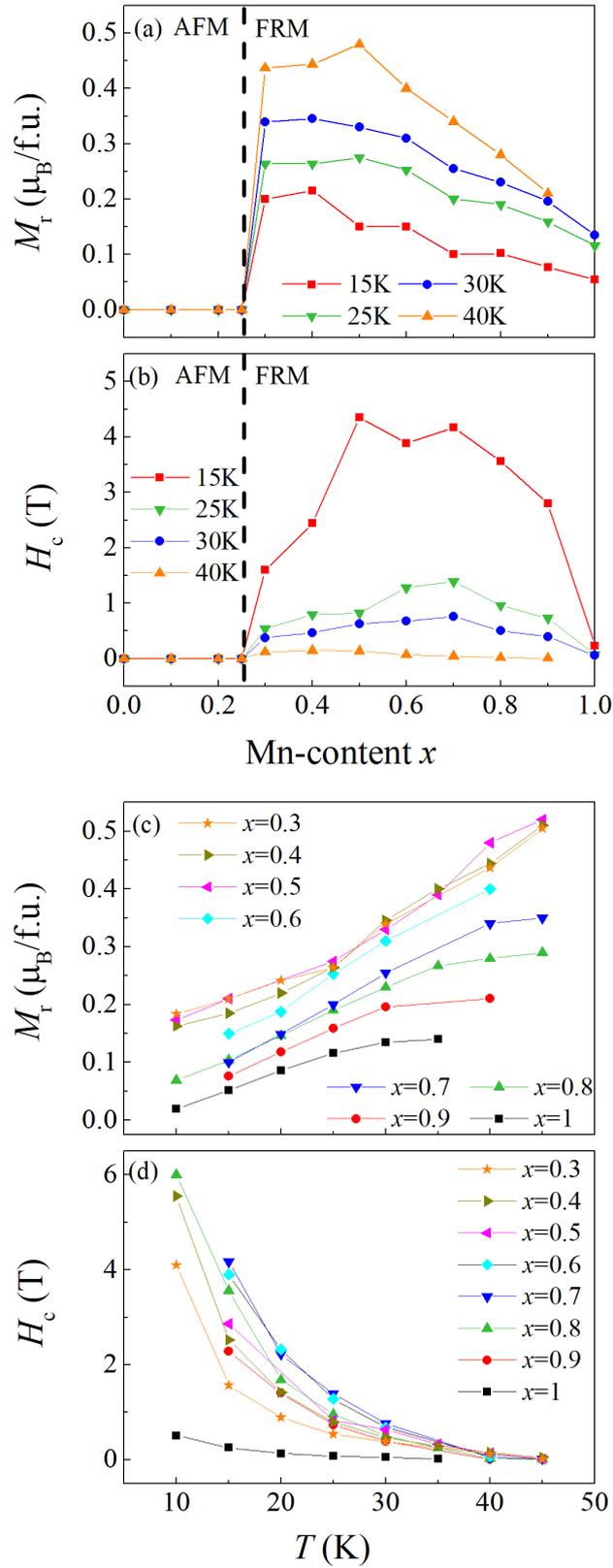

Figure 5. (a) and (b) present obtained remnant magnetization $M_r$ and coercive field $H_c$ at a function of $x$ at various temperatures, respectively. (c) and (d) show $T$-dependence of $M_r$ and $H_c$, respectively.

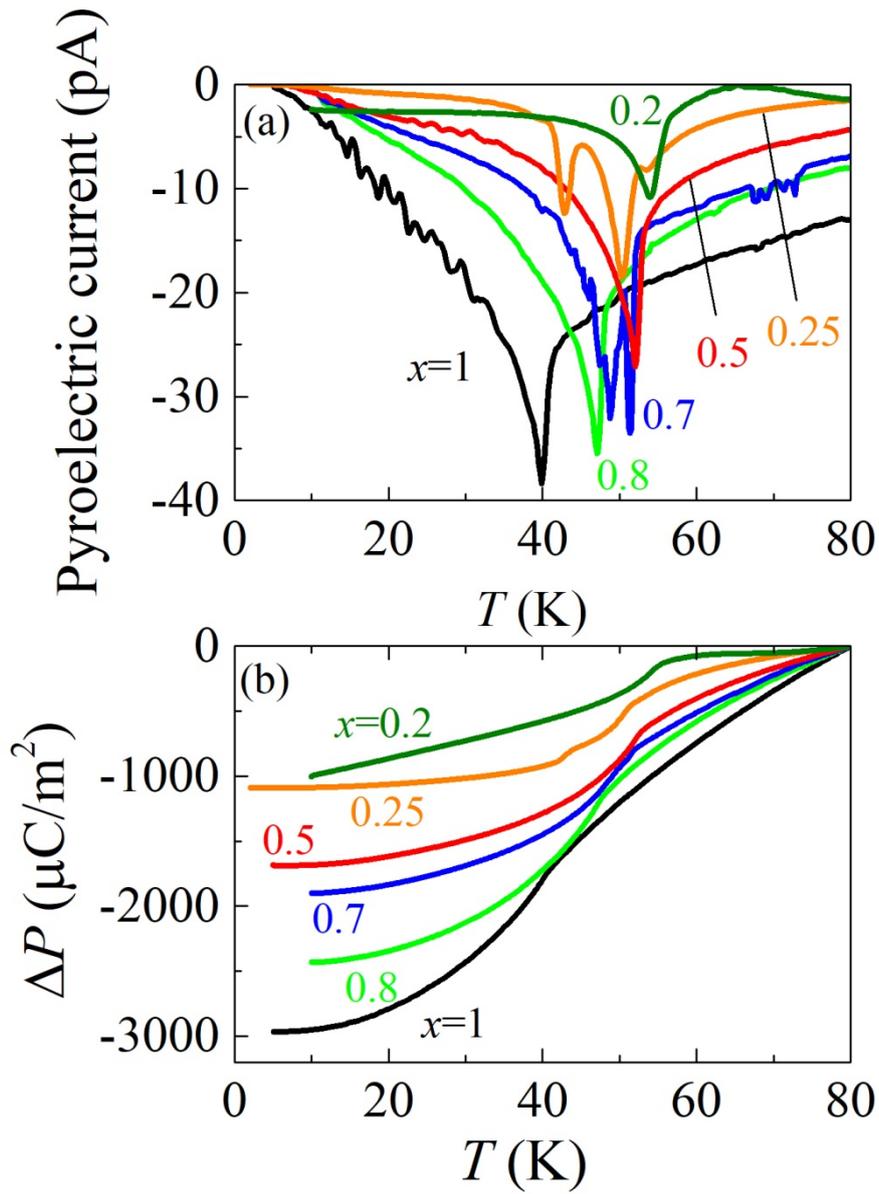

Figure 6. (a) Measured pyroelectric current as a function of $T$. (b) Integrated electric polarization as a function of $T$.

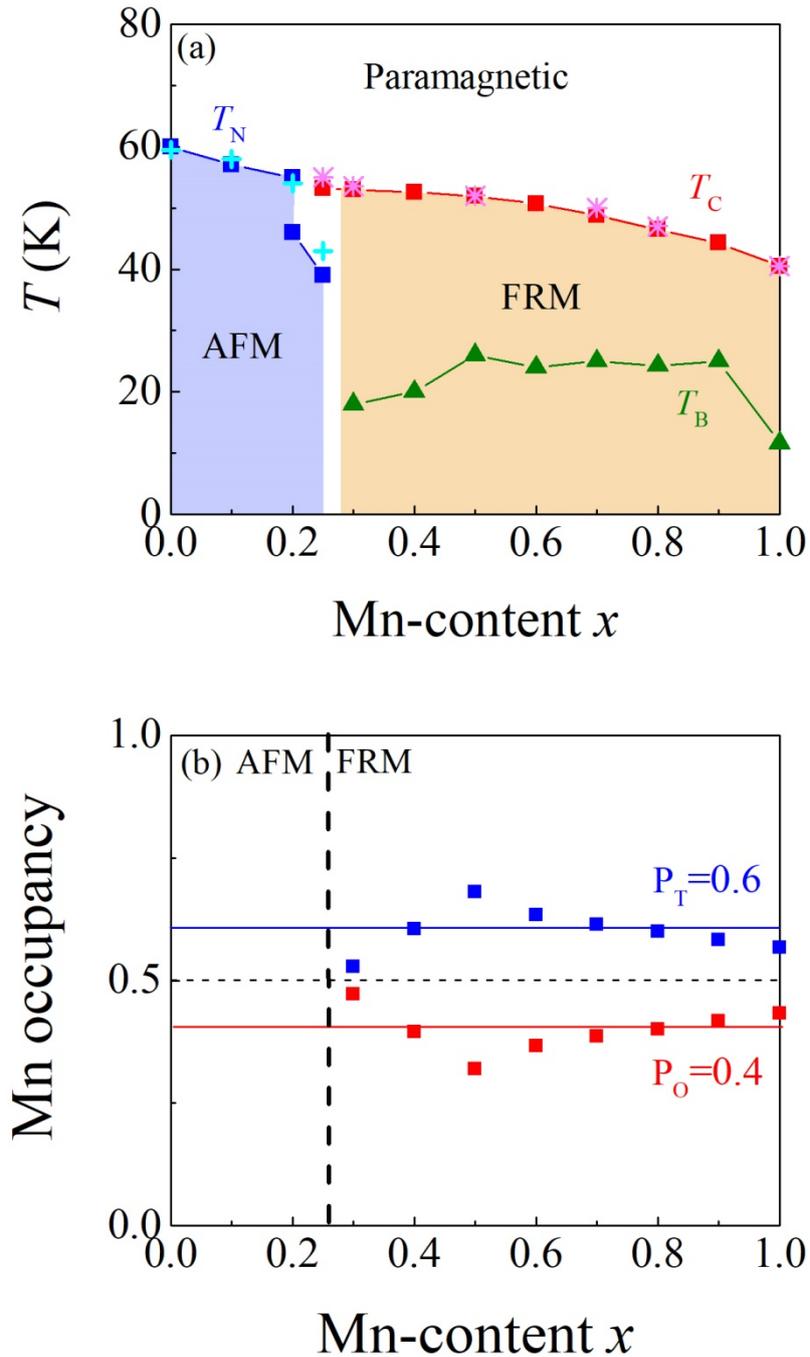

Figure 7. (a) *T-x* phase diagram is summarized for $(Fe_{1-x}Mn_x)_2Mo_3O_8$ ($0 \leq x \leq 1$), based on the data of magnetization (solid squares and triangles) and pyroelectric current (stars and crosses). (b) Calculated $Mn^{2+}$ occupation ratios $P_O$ and $P_T$ as a function of *x*.

**Table of Contents Graphic**

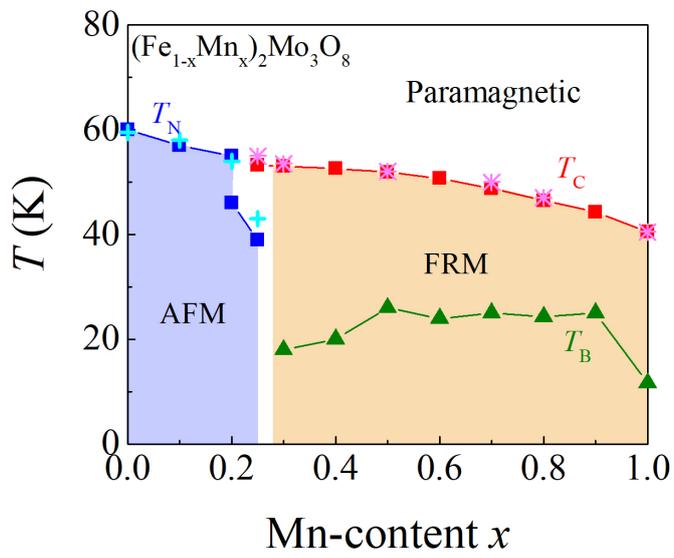
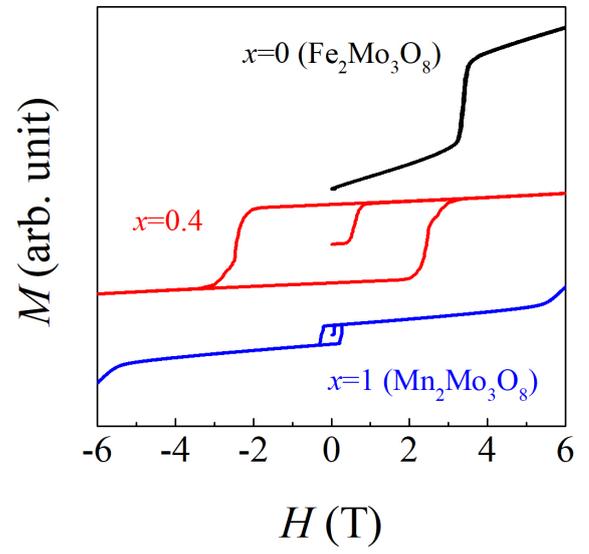